\documentclass[superscriptaddress,nofootinbib,notitlepage]{revtex4-1}
\usepackage{amssymb}
\usepackage{amsfonts}
\usepackage{amsmath}
\usepackage{graphicx}
\usepackage{dcolumn}
\usepackage{bm}
\usepackage[latin1]{inputenc}
\usepackage[dvips]{hyperref}

\begin{document}

\title{Electrovacuum geometries in five dimensions}
\author{Rafael Ferraro}\email[Member of Carrera del Investigador
Cient\'{\i}fico (CONICET, Argentina);
]{ferraro@iafe.uba.ar}\affiliation{Instituto de Astronom\'\i a y
F\'\i sica del Espacio (IAFE, CONICET-UBA), Casilla de Correo 67,
Sucursal 28, 1428 Buenos Aires, Argentina.}
\affiliation{Departamento de F\'\i sica, Facultad de Ciencias
Exactas y Naturales, Universidad de Buenos Aires, Ciudad
Universitaria, Pabell\'on I, 1428 Buenos Aires,
Argentina.\vskip.5cm}

\begin{abstract}
The Chong-Cveti\v{c}-L\"{u}-Pope $5D$ rotating charged black hole proves
to belong to a set of solutions to Einstein-Maxwell-Chern-Simons
(EMCS) equations that share the electromagnetic potential and the
Chern-Simons coupling constant but differ in the Kretschmann
invariant. This one-parametric family of solutions is found by
proposing a properly deformed Pleba\'{n}ski-Demia\'{n}ski Ansatz for
modeling the metric tensor. While no black-hole solutions for other
values of the Chern-Simons coupling constant are found within this
Ansatz, another type of non-static electrovacuum solutions to $5D$
EMCS equations are obtained, namely Kundt spacetimes sourced by a
pure-radiation field.
\end{abstract}

\maketitle

\section{Introduction}

\label{Intro} The search for solutions to Einstein equations can be
greatly facilitated if the solution is looked for in a family of
metrics conveniently prepared. For this aim, one uses an Ansatz
whose metrics are simple enough to make easy the control of the
Ricci tensor; at the same time they should be sufficiently flexible
to have a chance of finding solutions. A well known
example of such strategy is the Kerr-Schild (KS) Ansatz \cite%
{Kerr63,Kerr65,Debney67}, where the metric is written in the way ${g}_{\mu
\nu }=\overset{o}{g}_{\mu \nu }+f(x)\,k_{\mu }\,k_{\nu }$; here $\overset{o}{%
g}_{\mu \nu }$ is a \textquotedblleft seed\textquotedblright\ known metric, $%
k_{\mu }$ is a conveniently chosen null congruence of both metrics $\overset{%
o}{g}_{\mu \nu }$ and ${g}_{\mu \nu }$, and $f(x)$ is a free function
spanning the members of the family. Also Pleba\'{n}ski-Demia\'{n}ski (PD)
metrics \cite{PD} have proved to be a fruitful way of representing the host
family of metrics. The components of PD metrics depend on two coordinates $r$%
, $p$; two functions $X(p)$, $Y(r)$ play the role of free degrees of
freedom; besides they contain several parameters to be identified
with the cosmological constant, the angular momenta of the solution,
etc. In four dimensions, both KS and PD Ans\"{a}tze succeed in leading
to rotating black-hole solutions. Even the charged Kerr-Newman
solution belongs to both Ans\"{a}tze. Actually PD metrics in four
dimensions contain the most general axially symmetric solution to
\textquotedblleft source-less\textquotedblright\ Einstein-Maxwell
equations. This Petrov type-D solution is characterized by seven
parameters: the mass, the Newman-Unti-Tamburino (NUT) charge, the
angular momentum, the electric charge, the magnetic charge, the
acceleration, and the cosmological constant (see also \cite{PHG}).
For five and higher dimensions there exists a complete catalog of
axially symmetric solutions to the vacuum Einstein equations
\cite{MP,GLPP,CGLP}. Instead, the only charged rotating black hole
so far obtained is the $5D$ Chong-Cveti\v{c}-L\"{u}-Pope (CCLP) geometry \cite%
{CCLP}, which is a solution to Einstein-Maxwell-Chern-Simons (EMCS)
equations for a specific Chern-Simons (CS) coupling constant; it was
obtained in the context of five-dimensional minimal gauged
supergravity. The CCLP geometry is characterized by the mass, the
NUT charge, two angular momenta, the electric charge, and the
cosmological constant. There are not clues to build $5D$ solutions
to EMCS equations for other values of the CS coupling constant.
However there exist some attempts for the case of equal angular
momenta, which resort to numerical \cite{KN, BKNR} or perturbative
\cite{MM} techniques. Besides, a sector of the $5D$ Einstein-Maxwell
equations has proven to be integrable for a restricted form of the
electromagnetic field \cite{Y}.

\bigskip

We aim to develop a method to search for solutions to $5D$ EMCS
equations, by starting from a proper extension of the $5D$ PD Ansatz
\cite{CLP}. As a result, we will find that the CCLP solution appears
as one among other solutions to EMCS equations for the same
electromagnetic potential and CS coupling constant, but differing in
the value of the Kretschmann invariant. On the other hand, we will
also find a family of $5D$ geometries conformal to $pp$waves that
are sourced by a pure-radiation electromagnetic field. In Section
\ref{5D PD metrics} we will display the $5D$ PD Ansatz we will use
for introducing the host family of metrics. For these metrics, we
will show the eigenvalue-eigenvector structure of the Einstein
tensor, and their double Kerr-Schild form. In Section \ref{5D
Maxwell} we will compute the eigenvalue-eigenvector structure of the
energy-momentum tensor belonging to a rotating \textquotedblleft
pointlike\textquotedblright\ charge. Since the
eigenvalue-eigenvector structures of both Einstein and
energy-momentum tensors cannot be matched in the PD Ansatz, in
Section \ref{extendedPD} we will introduce an extended PD Ansatz. We
will obtain a one-parametric family of solutions accomplishing the
EMCS equations for the same electromagnetic potential and
Chern-Simons coupling constant $2\sqrt{G/3}$, among which the CCLP
solution is found. In Section \ref{further extensions} we will
discuss further extensions of the Ansatz; however we will not
succeed in getting black-hole solutions for other values of the
Chern-Simons coupling constant. In Section \ref{Pure radiation} we
will show a different type of non-static solutions to EMCS
equations, which is associated with pure-radiation sources. In
Section \ref{Conclusions} we will display the conclusions.

\section{Pleba\'{n}ski-Demia\'{n}ski metrics in five dimensions}

\label{5D PD metrics} We will look for rotating charged black-hole
solutions in the set of $5D$ Pleba\'{n}ski-Demia\'{n}ski-like
metrics of the form \cite{CLP}

\begin{equation}
\mathbf{g}~=-\frac{Y(r)}{p^{2}+r^{2}}~~\mathbf{\omega }^{0}\otimes \mathbf{%
\omega }^{0}+\frac{X(p)}{p^{2}+r^{2}}~~\mathbf{\omega }^{1}\otimes \mathbf{%
\omega }^{1}+\frac{a^{2}b^{2}}{p^{2}r^{2}}~~\mathbf{\omega }^{2}\otimes
\mathbf{\omega }^{2}+\left( p^{2}+r^{2}\right) ~\left( \frac{dr\otimes dr}{%
Y(r)}+\frac{dp\otimes dp}{X(p)}\right)  \label{PD}
\end{equation}

\noindent where

\begin{eqnarray}
\mathbf{\omega }^{0} &\equiv &\frac{(1-p^{2}\lambda )~dt}{(1-a^{2}~\lambda
)(1-b^{2}~\lambda )}-\frac{a~(a^{2}-p^{2})~d\phi }{(a^{2}-b^{2})(1-a^{2}~%
\lambda )}-\frac{b~(b^{2}-p^{2})~d\psi }{(b^{2}-a^{2})(1-b^{2}~\lambda )}~,
\notag \\
&&  \notag \\
&&  \notag \\
\mathbf{\omega }^{1} &\equiv &\frac{(1+r^{2}\lambda )~dt}{(1-a^{2}~\lambda
)(1-b^{2}~\lambda )}-\frac{a~(a^{2}+r^{2})~d\phi }{(a^{2}-b^{2})(1-a^{2}~%
\lambda )}-\frac{b~(b^{2}+r^{2})~d\psi }{(b^{2}-a^{2})(1-b^{2}~\lambda )}~,
\notag \\
&&  \notag \\
&&  \notag \\
\mathbf{\omega }^{2} &\equiv &-\frac{(1+r^{2}\lambda )~(1-p^{2}\lambda )~dt}{%
(1-a^{2}~\lambda )(1-b^{2}~\lambda )}+\frac{(a^{2}+r^{2})(a^{2}-p^{2})~d\phi
}{a~(a^{2}-b^{2})(1-a^{2}~\lambda )}+\frac{(b^{2}+r^{2})(b^{2}-p^{2})~d\psi
}{b~(b^{2}-a^{2})(1-b^{2}~\lambda )}~.  \label{omegas}
\end{eqnarray}
\bigskip

\noindent Here, $a$, $b$, $\lambda $ are three parameters which can be
freely chosen (however, it must be $|a|\neq |b|$).

\bigskip

Even though the $5D$ metrics (\ref{PD}) are deprived of some
features of $4D$ Pleba\'{n}ski-Demia\'{n}ski metrics (for instance,
they do not contain the parameter associated with the acceleration),
the Ansatz (\ref{PD}) should be enough for our purposes. The inverse
metric for this Ansatz is

\begin{equation}
\mathbf{g}^{-1}=-\frac{\mathbf{v}_{0}\otimes \mathbf{v}_{0}}{\left(
p^{2}+r^{2}\right) ~Y(r)}~+~\frac{\mathbf{v}_{1}\otimes \mathbf{v}_{1}}{%
\left( p^{2}+r^{2}\right) ~X(p)}~+~\frac{\mathbf{v}_{2}\otimes \mathbf{v}_{2}%
}{p^{2}~r^{2}}~+\frac{Y(r)}{p^{2}+r^{2}}~\frac{\partial }{\partial r}\otimes
\frac{\partial }{\partial r}~+~\frac{X(p)}{p^{2}+r^{2}}~\frac{\partial }{%
\partial p}\otimes \frac{\partial }{\partial p}~, \bigskip\bigskip
\label{IPD}
\end{equation}
\noindent where

\begin{eqnarray}
\mathbf{v}_{0} &\equiv &r^{-2}~(a^{2}+r^{2})~(b^{2}+r^{2})~\left( \frac{%
\partial }{\partial t}+\frac{a~(1+r^{2}\lambda )}{a^{2}+r^{2}}~\frac{%
\partial }{\partial \phi }+\frac{b~(1+r^{2}\lambda )}{b^{2}+r^{2}}~\frac{%
\partial }{\partial \psi }\right) ~,  \notag \\
&&  \notag \\
&&  \notag \\
\mathbf{v}_{1} &\equiv &p^{-2}~(a^{2}-p^{2})~(b^{2}-p^{2})~\left( \frac{%
\partial }{\partial t}+\frac{~a~(1-p^{2}\lambda )}{a^{2}-p^{2}}~\frac{%
\partial }{\partial \phi }+\frac{b~(1-p^{2}\lambda )}{b^{2}-p^{2}}~\frac{%
\partial }{\partial \psi }\right) ~,  \notag \\
&&  \notag \\
&&  \notag \\
\mathbf{v}_{2} &\equiv &a~b~\left( \frac{\partial }{\partial t}+\frac{1}{a}~%
\frac{\partial }{\partial \phi }+\frac{1}{b}~\frac{\partial }{\partial \psi }%
\right) ~.
\end{eqnarray}%
\vskip1cm

\noindent The orthogonal bases $\mathbf{\omega }^{i}=\{\mathbf{\omega }^{0},%
\mathbf{\omega }^{1},\mathbf{\omega }^{2},dr,dp\} $ and $\mathbf{v}_{i}=\{%
\mathbf{v}_{0},\mathbf{v}_{1},\mathbf{v}_{2},\partial /\partial r,\partial
/\partial p\}$ are dual, except for normalization factors. In fact it is
\begin{equation}
\mathbf{\omega }^{i}(\mathbf{v}_{j})=0~,~~~~\ \ \ ~\ ~\forall i\neq
j~.\bigskip
\end{equation}
They can be normalized:

\begin{eqnarray}
\mathbf{\omega }^{\widehat{0}}=\sqrt{\frac{Y(r)}{p^{2}+r^{2}}}\ \mathbf{%
\omega }^{0}~,~~~~\mathbf{\omega }^{\widehat{1}}=\sqrt{\frac{X(p)}{%
p^{2}+r^{2}}}\ \mathbf{\omega }^{1}~,~~~~\mathbf{\omega }^{\widehat{2}}=%
\frac{a~b}{p~r}\ \mathbf{\omega }^{2}~,~~~~\mathbf{\omega }^{\widehat{3}} =%
\sqrt{\frac{p^{2}+r^{2}}{Y(r)}}\ dr~,~~~~\mathbf{\omega }^{\widehat{4}} =%
\sqrt{\frac{p^{2}+r^{2}}{X(p)}}\ dp~,~~~~  \notag \\
\notag \\
\notag \\
\mathbf{v}_{\widehat{0}} =\frac{\mathbf{v}_{0}}{~\sqrt{( p^{2}+r^{2})Y(r)}}%
~,~~\mathbf{v}_{\widehat{1}}=\frac{\mathbf{v}_{1}}{~\sqrt{(p^{2}+r^{2}) X(p)}%
}~,~~\mathbf{v}_{\widehat{2}}=\frac{\mathbf{v}_{2}}{~p~r}~, ~~\mathbf{v}_{%
\widehat{3}}=\sqrt{\frac{Y(r)}{p^{2}+r^{2}}}\ \frac{\partial}{\partial r}~,~~%
\mathbf{v}_{\widehat{4}}=\sqrt{\frac{X(p)}{p^{2}+r^{2}}}\ \frac{\partial}{%
\partial p}~.~~~~ \bigskip\bigskip\bigskip  \label{normal}
\end{eqnarray}

\vskip.5cm

Some of the characteristics of the $5D$ Pleba\'{n}ski-Demia\'{n}ski metrics (%
\ref{PD}) are as follows,

\bigskip

i) They are invariant under the change $r^{2}\longleftrightarrow -p^{2}$, $%
X\longleftrightarrow Y$.

\bigskip

ii) The metric is Lorentzian only if $X>0$ and $Y>0$, or $X>0$ and $Y<0$
(however $\partial /\partial r$ would be timelike in the second case).

\bigskip

iii) $\left( p^{2}+r^{2}\right) \mathbf{g}^{-1}$ separates into terms
depending only on $r$ or $p$. In particular, this implies that
Hamilton-Jacobi equation is separable.

\bigskip

iv) In the chart $(t,\phi ,\psi ,r,p)$ the determinant is%
\begin{equation}
g=\det [\mathbf{g}]=-\frac{p^{2}r^{2}~\left( p^{2}+r^{2}\right) ^{2}}{%
(a^{2}-b^{2})^{2}(1-a^{2}~\lambda )^{2}(1-b^{2}~\lambda )^{2}}~.  \label{det}
\end{equation}

\bigskip

v) Einstein tensor $G^{\mu }_{\,\nu }$ is linear in the functions $X$, $Y$.
In five dimensions only the first and second derivatives of $X$ and $Y$
appear in $G^{\mu }_{\,\nu }$. Because of this reason the solutions $X$, $Y$
to vacuum Einstein equations come with additive free constants (see item
vii).\footnote{%
This property could be also obtained in four dimensions by properly
redefining $X$, $Y$.}

\bigskip

vi) $\left( p^{2}+r^{2}\right)\,R$ exhibits separation of variables:%
\begin{equation}
\left( p^{2}+r^{2}\right)\,R~=-p^{-2}~\left[ p^{2}~X^{\prime
}(p)-2~a^{2}~b^{2}~p^{-1}\right] ^{\prime }-r^{-2}~\left[ r^{2}~Y^{\prime
}(r)+2~a^{2}~b^{2}~r^{-1}\right] ^{\prime }~.
\end{equation}

\bigskip

vii) The solutions to vacuum Einstein equations are
\begin{equation}
X_{vac}(p)=-p^{-2}(a^{2}-p^{2})(b^{2}-p^{2})(1+p^{2}\lambda )+\alpha
~p^{2}+2~n~,  \label{Xvac}
\end{equation}%
\begin{equation}
Y_{vac}(r)=r^{-2}(a^{2}+r^{2})(b^{2}+r^{2})(1-r^{2}\lambda )-\alpha
~r^{2}-2~m~,  \label{Yvac}
\end{equation}%
where $\alpha $, $m$, $n$ are integration constants. This is the Kerr-NUT
solution with mass $m$, NUT charge $n$, angular momenta $a$, $b$, and
cosmological constant $\Lambda =6\lambda $ (however, see item viii) \cite%
{GLPP,CGLP,DKL,CLP}. The integration constant $\alpha $ represents a
choice of chart for the Kerr-NUT geometry within the PD Ansatz. The
Kretschmann invariant is $R_{~~\lambda \rho }^{\mu \nu }~R_{~~\mu
\nu }^{\lambda \rho }|_{vac}=40~\lambda ^{2}+96~(m+n)^{2}\left(
p^{2}+r^{2}\right) ^{-6}(3~p^{4}-10~p^{2}r^{2}+3~r^{4})$. Differing
from $4D$, in $5D$ the mass $m$ and the NUT charge $n$ are not
constrained to vanish in the de Sitter
geometry: it is enough that $n=-m$.\footnote{%
In five dimensions, $R_{\mu \nu \lambda \rho }-\frac{\Lambda }{6}~(g_{\mu
\lambda }~g_{\nu \rho }-g_{\mu \rho }~g_{\nu \lambda })$ is affected by a
global factor $(m+n)$. Actually, as shown in Ref.~\cite{CLP}, for odd
dimensions there exists a scaling symmetry making trivial one of the free
parameters. Thus, the NUT parameter $n$ could be removed by properly
changing coordinates, and the mass $m$ would be completely fixed in the de
Sitter geometry.}

\bigskip

viii) To guarantee the positiveness of $X_{vac}$, $Y_{vac}$ for the entire
range of the coordinates, we can choose $\alpha =0$, and replace $p$ with
the coordinate $\theta $ defined as%
\begin{equation}
p^{2}=a^{2}\cos ^{2}\theta +b^{2}\sin ^{2}\theta ~,  \label{p-theta}
\end{equation}%
i.e.,%
\begin{equation}
\cos ^{2}\theta =\frac{b^{2}-p^{2}}{b^{2}-a^{2}}~,~~~~~~~~~~~~\sin
^{2}\theta =\frac{a^{2}-p^{2}}{a^{2}-b^{2}}~.  \label{theta-p}
\end{equation}%
By replacing (\ref{p-theta}) and (\ref{theta-p}) in the Eq.~(\ref{Xvac}),
one sees that $X_{vac}(\theta )$ will be positive definite if $\lambda $, $n$
are properly chosen. Besides it is
\begin{equation}
dp^{2}=\frac{(b^{2}-a^{2})^{2}~\sin ^{2}\theta ~\cos ^{2}\theta }{a^{2}\cos
^{2}\theta +b^{2}\sin ^{2}\theta }~d\theta ^{2}~.
\end{equation}

\bigskip

ix) $\{\mathbf{v}_{i}\}=\{\mathbf{v}_{0},\mathbf{v}_{1},\mathbf{v}%
_{2},\partial /\partial r,\partial /\partial p\}$ are eigenvectors of the
Ricci tensor. This is a very important property of the Pleba\'{n}ski-Demia%
\'{n}ski Ansatz, since it implies that the structure of the Einstein
tensor
is known independently of the functions $X$, $Y$. Indeed the functions $X$, $%
Y$ only have to do with the eigenvalues of the tensor. Without loss
of generality we can write
\begin{equation}
X(p)=X_{vac}(p)+f(p)~,  \label{XPDansatz}
\end{equation}%
\begin{equation}
Y(r)=Y_{vac}(r)+g(r)~,  \label{YPDansatz}
\end{equation}%
\begin{equation}
(G_{\,\nu }^{\mu }+\Lambda ~\delta _{~\nu }^{\mu })~v_{i}^{\nu }=\epsilon
_{i}~v_{i}^{\nu }~.
\end{equation}%
Thus $\mathbf{v}_{0}$ and $\partial /\partial r$ share the eigenvalue%
\begin{equation}
\epsilon _{0}=\frac{2~r^{3}~f^{\prime }+p~(p^{2}+3~r^{2})~g^{\prime }}{%
2~p~r~(p^{2}+r^{2})^{2}}+\frac{f^{\prime \prime }}{2~(p^{2}+r^{2})}~,
\label{eg0}
\end{equation}%
and $\mathbf{v}_{1}$ and $\partial /\partial p$ share the eigenvalue%
\begin{equation}
\epsilon _{1}=\frac{2~p^{3}~g^{\prime }+r~(r^{2}+3~p^{2})~f^{\prime }}{%
2~p~r~(p^{2}+r^{2})^{2}}+\frac{g^{\prime \prime }}{2~(p^{2}+r^{2})}~;
\label{eg1}
\end{equation}%
therefore it results%
\begin{equation}
\epsilon _{0}+\epsilon _{1}=\frac{p^{-3}~(p^{3}~f^{\prime })^{\prime
}+r^{-3}~(r^{3}~g^{\prime })^{\prime }}{2~(p^{2}+r^{2})}~.  \label{eg0+eg1}
\end{equation}
Besides $\mathbf{v}_{2}$ has the eigenvalue%
\begin{equation}
\epsilon _{2}=\frac{f^{\prime \prime }+g^{\prime \prime }}{2~(p^{2}+r^{2})}%
~. \bigskip  \label{eg2}
\end{equation}

In terms of the normalized bases (\ref{normal}), it follows that
\begin{equation}
(G_{~\nu }^{\mu }+\Lambda ~\delta _{~\nu }^{\mu })~\frac{\partial }{\partial
x^{\mu }}\otimes dx^{\nu }=\epsilon _{0}~\left( ~\mathbf{v}_{\widehat{0}%
}\otimes \mathbf{\omega }^{\widehat{0}}+\frac{\partial }{\partial r}\otimes
dr\right) +\epsilon _{1}~\left( ~\mathbf{v}_{\widehat{1}}\otimes \mathbf{%
\omega }^{\widehat{1}}+\frac{\partial }{\partial p}\otimes dp\right)
+\epsilon _{2}~~\mathbf{v}_{\widehat{2}}\otimes \mathbf{\omega }^{\widehat{2}%
}~.  \label{Einstein}
\end{equation}%
While this result can be considered as the strength of Pleba\'{n}ski-Demia%
\'{n}ski approach, at the same time it is its weakness. In fact, the
expression (\ref{Einstein}) constitutes a severe limitation for the Pleba%
\'{n}ski-Demia\'{n}ski Ansatz as a way to generate solutions to
Einstein equations with sources: the equations should be sourced by
an energy-momentum tensor with the same eigenvalue-eigenvector
structure.

\bigskip

x) Pleba\'{n}ski-Demia\'{n}ski metrics can be cast in a Kerr-Schild form
that is linear in the function $Y(r)$. Those contributions being inversely
proportional to $Y(r)$ are absorbed into a coordinate change. Even more, by
allowing for complex coordinates, one can also obtain a double Kerr-Schild
form that is linear in $X$, $Y$. In fact, let us perform the complex
coordinate change
\begin{eqnarray}
~dt^{\prime } &=&dt+\frac{(a^{2}+r^{2})(b^{2}+r^{2})}{r^{2}~Y(r)}~dr+i~\frac{%
(a^{2}-p^{2})(b^{2}-p^{2})}{p^{2}~X(p)}~dp~,~~~~~~~  \notag \\
d\phi ^{\prime } &=&d\phi -\lambda ~a~dt^{\prime }+\frac{%
a~(b^{2}+r^{2})(1+r^{2}\lambda )}{r^{2}~Y(r)}~dr+i~\frac{%
a~(b^{2}-p^{2})(1-p^{2}\lambda )}{p^{2}~X(p)}~dp~,  \notag \\
d\psi ^{\prime } &=&d\psi -\lambda ~b~dt^{\prime }+\frac{%
b~(a^{2}+r^{2})(1+r^{2}\lambda )}{r^{2}~Y(r)}~dr+i~\frac{%
b~(a^{2}-p^{2})(1-p^{2}\lambda )}{p^{2}~X(p)}~dp~,  \notag \\
dr^{\prime } &=&dr~,~~~~~~~~~~~~~~~dp^{\prime }=i~dp~.\bigskip \bigskip
\end{eqnarray}%
In the new coordinate basis, the vectors $\mathbf{v}_{i}$'s look
like
\begin{eqnarray}
\mathbf{v}_{0} &=&r^{-2}~(a^{2}+r^{2})~(b^{2}+r^{2})~\left( \frac{\partial }{%
\partial t^{\prime }}+\frac{a~(1-a^{2}\lambda )}{a^{2}+r^{2}}~\frac{\partial
}{\partial \phi ^{\prime }}+\frac{b~(1-b^{2}\lambda )}{b^{2}+r^{2}}~\frac{%
\partial }{\partial \psi ^{\prime }}\right) ~,  \notag \\
&&  \notag \\
\mathbf{v}_{1} &=&p^{-2}~(a^{2}-p^{2})~(b^{2}-p^{2})~\left( \frac{\partial }{%
\partial t^{\prime }}+\frac{a~(1-a^{2}\lambda )}{a^{2}-p^{2}}~\frac{\partial
}{\partial \phi ^{\prime }}+\frac{b~(1-b^{2}\lambda )}{b^{2}-p^{2}}~\frac{%
\partial }{\partial \psi ^{\prime }}\right) ~,  \notag \\
&&  \notag \\
\mathbf{v}_{2} &=&a~b~\left( \frac{\partial }{\partial t^{\prime }}%
+a^{-1}~(1-a^{2}\lambda )~\frac{\partial }{\partial \phi ^{\prime }}%
+b^{-1}~(1-b^{2}\lambda )~\frac{\partial }{\partial \psi ^{\prime }}\right)
~.
\end{eqnarray}%
Besides it is
\begin{eqnarray}
\frac{\partial }{\partial r}~ &=&~\frac{\partial }{\partial r^{\prime }}+%
\frac{\mathbf{v}_{0}}{Y(r)}~,  \label{d/dr} \\
&&  \notag \\
~\frac{\partial }{\partial p}~ &=&~i~\frac{\partial }{\partial p^{\prime }}%
+i~\frac{\mathbf{v}_{1}}{X(p)}~.  \label{d/dp}
\end{eqnarray}%
Thus, the inverse metric (\ref{IPD}) is equal to
\begin{equation}
\mathbf{g}^{-1}=\frac{1}{p^{2}+r^{2}}\left[\frac{\partial }{\partial
r^{\prime }}\otimes \mathbf{v}_{0}+\mathbf{v}_{0}\otimes \frac{\partial }{%
\partial r^{\prime }}-\frac{\partial }{\partial p^{\prime }}\otimes \mathbf{v%
}_{1}-\mathbf{v}_{1}\otimes \frac{\partial }{\partial p^{\prime }}\right]+~
\frac{\mathbf{v}_{2}\otimes \mathbf{v}_{2}}{p^{2}~r^{2}}~+~\frac{Y(r)}{%
p^{2}+r^{2}}~\frac{\partial }{\partial r^{\prime }}\otimes \frac{\partial }{%
\partial r^{\prime }}~-~\frac{X(p)}{p^{2}+r^{2}}~\frac{\partial }{\partial
p^{\prime }}\otimes \frac{\partial }{\partial p^{\prime }}~;
\label{5Dlinearmetric}
\end{equation}

\noindent so in the complex chart $(t^{\prime },\phi ^{\prime },\psi
^{\prime },r^{\prime },p^{\prime })$ their components are linear in $X$, $Y$%
. The vectors%
\begin{equation}
\mathbf{k~~}\equiv ~-\frac{\partial }{\partial r^{\prime }}=-\frac{\partial
}{\partial r}+\frac{\mathbf{v}_{0}}{Y(r)}~,  \label{k}
\end{equation}%
\begin{equation}
\mathbf{K~~}\equiv ~-\frac{\partial }{\partial p^{\prime }}=i~\frac{\partial
}{\partial p}+\frac{\mathbf{v}_{1}}{X(p)}~,  \label{K}
\end{equation}%
are null and geodesic whatever the functions $X$, $Y$ are; besides
they are mutually perpendicular. Therefore, the metric
(\ref{5Dlinearmetric}) has a double Kerr-Schild form, with functions
$X$, $Y$ playing the role of free degrees of freedom in the
Kerr-Schild Ansatz (however, in this case they are restricted to
depend on a unique coordinate). In particular, the Kerr-NUT
metric --whose functions $X$, $Y$ are given in Eqs.~(\ref{Xvac}) and (\ref%
{Yvac})-- can be cast into this form:\footnote{%
As mentioned in Property (vii), $m=0=n$ is just a particular case of de
Sitter geometry. In five dimensions, the geometry is still de Sitter
whenever $n$ is equal and opposite to $m$.}
\begin{equation}
\mathbf{g}_{Kerr-NUT}^{-1}~=~\mathbf{g}_{dS}^{-1}~-~\frac{2~m}{p^{2}+r^{2}}~%
\mathbf{k}\otimes \mathbf{k}-~\frac{2~n}{p^{2}+r^{2}}\mathbf{~K\otimes K~~.}
\end{equation}

\section{\protect\bigskip Einstein-Maxwell equations}

\label{5D Maxwell} Both the Kerr-Schild and
Pleba\'{n}ski-Demia\'{n}ski Ans\"{a}tze have interesting properties
regarding Maxwell equations. Let us explain them in the context of
the KS Ansatz; the conclusions will be also
valid for the double KS form exhibited by PD metrics. Let be the metric \cite%
{Kerr63,Kerr65,Debney67}
\begin{equation}
{g}_{\mu \nu }\ =\ \overset{o}{g}_{\mu \nu }+f(x)\,k_{\mu }\,k_{\nu }~
\label{KSansatz}
\end{equation}%
where $k_{\mu }$ is a null vector of $\overset{o}{g}_{\mu \nu }$. Then, the
determinant $g=\det ({g}_{\mu \nu })$ does not depend on $f(x)$, and $k_{\mu
}$ is a null vector of ${g}_{\mu \nu }$ too. Besides, the inverse metric
reads
\begin{equation}
{g}^{\mu \nu }\ =\ \overset{o}{g}\,^{\mu \nu }-f(x)\,k^{\mu }\,k^{\nu }\,.
\end{equation}%
If $k_{\mu }$ is not only null but geodesic too, i.e.%
\begin{equation}
k^{\mu }\,k_{\mu }\ =\ 0\,~~~\ \ ~~~~~~\text{and\thinspace }~~~\ \ ~~~~~~0\
=\ k^{\mu }\,(k_{\nu ;\mu }-k_{\mu ;\nu })\ =\ k^{\mu }\,(\partial _{\mu
}k_{\nu }-\partial _{\nu }k_{\mu })~,
\end{equation}%
where $k^{\mu }=\overset{o}{g}\,^{\mu \nu }k_{\nu }={g}^{\mu \nu }k_{\nu }$,
then it can be proved that if an electromagnetic potential $A_{\mu
}=A(x)\,k_{\mu }$ solves Maxwell equations in the metric ${\overset{o}{g}}%
_{\mu \nu }$ then it will also solve them in the metric ${g}_{\mu \nu }$. In
fact, it is easy to prove that the field tensor
\begin{equation}
F^{\mu \nu }\ =\ {g}^{\mu \lambda }\,{g}^{\nu \rho }\,(\partial _{\lambda
}A_{\rho }-\partial _{\rho }A_{\lambda })
\end{equation}%
does not depend on the function $f(x)$; so the equations%
\begin{equation}
\partial _{\mu }\left( \sqrt{-g}~F^{\mu \nu }\right) =0
\end{equation}%
are not affected by $f(x)$. Concerning the Einstein-Maxwell problem,
this property implies that the Einstein equations will be sourced by
an energy-momentum tensor, the electromagnetic energy-momentum
tensor
\begin{equation}
T_{\ \nu }^{\mu }\ =\ -\frac{1}{4\pi }\,\left( F^{\mu \rho }F_{\rho \nu }-%
\frac{1}{4}\,{\delta }_{\nu }^{\mu }\,F^{\lambda \rho }\,F_{\rho \lambda
}\right) \,  \label{tmunu}
\end{equation}%
(we use the signature $(-+++...)$), which does not contain the unknown
function $f(x)$. Moreover, as proved in References \cite{DG,GG,GLPP}, if $%
k^{\mu }$ is tangent to a (null-) geodesic congruence, then
$R_{~~\nu }^{\mu }$ will be linear in $f(x)\,k^{\mu }\,k_{\nu }$ .
So, in terms of the function $f(x)$ the problem gets rather simple,
since $f(x)$ appears only linearly in the Einstein tensor. Of
course, the success of the Kerr-Schild Ansatz cannot be a priori
guaranteed. To have a chance of finding a new solution to Einstein
or Einstein-Maxwell equations one should start from a suitable null
vector $k^{\mu }$ in order that the sole unknown function $f(x) $
can fulfill the entire set of equations.

In the framework of PD metrics, let us consider the potential of a rotating
\textquotedblleft pointlike\textquotedblright\ charge%
\begin{equation}
A_{\mu }=\frac{Q}{p^{2}+r^{2}}~k_{\mu }~\mathbf{,}  \label{potentialk}
\end{equation}%
where $k_{\mu }$ are the covariant components of the null vector (\ref{k}).
We remark that this potential is equivalent to $\mathbf{A}%
=Q~(p^{2}+r^{2})^{-1}~\mathbf{\omega }^{0}$ and $\mathbf{A}%
=Q~(p^{2}+r^{2})^{-1}~\mathbf{\omega }^{1}$, since they differ in pure gauge
terms.\footnote{%
Notice that $k_{\mu }~dx^{\mu }=-(p^{2}+r^{2})~Y(r)^{-1}~dr+~\mathbf{\omega }%
^{0}$. Besides, from definitions (\ref{omegas}), it is easy to verify that $%
(p^{2}+r^{2})^{-1}~(\mathbf{\omega }^{0}-\mathbf{\omega }^{1})$ is a
closed 1-form.\label{gaugeequivalence}} The field
$\mathbf{F}=d\mathbf{A}$ verifies
the Maxwell equations in the metric (\ref{PD}) whatever the functions $X(p)$%
, $Y(r)$ are. By computing the eigenvalues and eigenvectors of the
energy-momentum tensor (\ref{tmunu}) it follows that%
\begin{equation}
8\pi G~T_{~\nu }^{\mu }~\frac{\partial }{\partial x^{\mu }}\otimes dx^{\nu
}=-\frac{4~G~Q^{2}}{(p^{2}+r^{2})^{3}}~\left[ \mathbf{v}_{\widehat{0}%
}\otimes \mathbf{\omega }^{\widehat{0}}+\frac{\partial }{\partial r}\otimes
dr-~\mathbf{v}_{\widehat{1}}\otimes \mathbf{\omega }^{\widehat{1}}-\frac{%
\partial }{\partial p}\otimes dp+\frac{p^{2}-r^{2}}{p^{2}+r^{2}}~\mathbf{v}_{%
\widehat{2}}\otimes \mathbf{\omega }^{\widehat{2}}\right] ~  \label{source}
\end{equation}%
We will try to combine this result with the one in
Eq.~(\ref{Einstein}), to know whether there is a chance of getting a
solution to $5D$ Einstein-Maxwell equations within the Ansatz
(\ref{PD}), (\ref{XPDansatz}), and (\ref{YPDansatz}). To this aim,
we should find functions $f(p)$, $g(r)$ such that the eigenvalues
(\ref{eg0}-\ref{eg2}) become equal to the ones in the
energy-momentum tensor of Eq.~(\ref{source}). By comparing
Eqs.~(\ref{Einstein}) and (\ref{source}) we should search for the
eigenvalues%
\begin{equation}
\epsilon _{0}=-\epsilon _{1}=-\frac{4~G~Q^{2}}{(p^{2}+r^{2})^{3}}~,~~~~~\
~~~~~~~~\epsilon _{2}=-\frac{4~G~Q^{2}(p^{2}-r^{2})}{(p^{2}+r^{2})^{4}}~.
\label{1}
\end{equation}%
The condition $\epsilon _{0}+\epsilon _{1}=0$, which results from the
expected eigenvalues (\ref{1}), implies $f^{\prime }=-B~p+C~p^{-3}$ and $%
g^{\prime }=B~r+D~r^{-3}$ in Eq.~(\ref{eg0+eg1}). However, these functions $%
f(p)$, $g(r)$ do not lead to the expected eigenvalues. Therefore,
there is not a solution sourced by the rotating \textquotedblleft
pointlike\textquotedblright\ charge potential (\ref{potentialk})
within the
Ansatz (\ref{PD}).\footnote{%
The static case is not included here because the Ansatz prevents the
simultaneous vanishing of $a$ and $b$ (see the determinant
(\ref{det})).}

\bigskip

\section{Extending the Pleba\'{n}ski-Demia\'{n}ski Ansatz}

\label{extendedPD}

As a way to enlarge the set of Pleba\'{n}ski-Demia\'{n}ski metrics,
and thus improve the chance of getting solutions to Einstein-Maxwell
equations, we could try including a \textquotedblleft
conformal\textquotedblright\ factor in the four-dimensional sector
where the electromagnetic field manifests
itself (notice that $F_{\mu \nu }~v_{2}^{\nu }=0$):\footnote{%
This kind of metric has been used in Ref.~\cite{LMP2} to obtain magnetic
dipole-charged solutions with electric charge in five-dimensional minimal
supergravity.}
\begin{equation}
\mathbf{g}~=\Theta (r,p)~\left[ -\frac{Y(r)}{p^{2}+r^{2}}~~\mathbf{\omega }%
^{0}\otimes \mathbf{\omega }^{0}+\frac{X(p)}{p^{2}+r^{2}}~~\mathbf{\omega }%
^{1}\otimes \mathbf{\omega }^{1}+\left( p^{2}+r^{2}\right) ~\left( \frac{%
dr\otimes dr}{Y(r)}+\frac{dp\otimes dp}{X(p)}\right) \right] +\frac{%
a^{2}b^{2}}{p^{2}r^{2}}~~\mathbf{\omega }^{2}\otimes \mathbf{\omega }^{2}~.
\label{conformal}
\end{equation}%
The potential (\ref{potentialk}) will still satisfy Maxwell equations, and
its energy-momentum tensor will scale with $\Theta (r,p)^{-2}$:%
\begin{equation}
T_{~\nu }^{\mu }\longrightarrow \Theta (r,p)^{-2}~T_{~\nu }^{\mu }~,
\label{conft}
\end{equation}%
which means that the so-modified energy-momentum tensor still
possesses the structure displayed in Eqs.~(\ref{source}). On the
other hand, to save the structure of $G_{~\nu }^{\mu }$ displayed in
Eq.~(\ref{Einstein}), we are
compelled to employ the function $\Theta (r,p)=(B~p^{2}r^{2}+C)^{-2}$ ($B$, $%
C$ are integration constants). However, it is not possible to find a set $B$%
, $C$, $X(p)$, $Y(r)$ matching the eigenvalues of the source (\ref{source})-(%
\ref{conft}).\footnote{%
Within this context, we remark the existence of a non-flat vacuum solution:
if $\lambda =0$, then the Ricci tensor is zero for $C=0$, $X_{vac}(p)=\alpha
\,p^{2}+\beta ~p^{4}-a^{2}b^{2}B^{2}p^{6}$, and $Y_{vac}(r)=-\alpha
\,r^{2}+\gamma ~r^{4}+a^{2}b^{2}B^{2}r^{6}$ ($\alpha $, $\beta $, $\gamma $
are integration constants). This solution is intrinsically curved, since the
Riemann tensor cannot be made zero by choosing the integration constants
(for other Ricci-flat solutions of this sort, see Ref.~\cite{LMP}). Instead,
its $4D$ analog --which is $\lambda =0$, $\Theta =(B\,p\,r)^{-2}$, $%
X_{vac}(p)=\alpha \,p^{2}+\beta \,p^{4}$, $Y_{vac}(r)=-\alpha \,r^{2}+\beta
\,r^{4}$-- is the flat spacetime for any values of the integration constants
$\alpha $, $\beta $.}

So, we should consider other ways of relaxing the Ansatz (\ref{PD})
in order that a source like (\ref{source}) makes sense in Einstein
equations. Let us remark that the source (\ref{source}) requires the
vanishing of $R_{~p}^{r}$. If the metric depends on only two
coordinates $r$, $p$, and it is diagonal in the block $(r,p)$, then
$R_{rp}$ has the form \cite{LL}
\begin{eqnarray}
R_{rp} &=&\frac{1}{2}~\left[ \log (-g^{-1}~g_{rr}~g_{pp})\right] _{,rp}
\label{Rrp} \\
&&-\frac{1}{4}~\left[ \log g_{rr}\right] _{,p}~~\left[ \log (-g^{-1}~g_{pp})%
\right] _{,r}-\frac{1}{4}~\left[ \log g_{pp}\right] _{,r}~~\left[ \log
(-g^{-1}~g_{rr})\right] _{,p}+\frac{1}{4}~g_{~~~,r}^{\mu \nu }~g_{\mu \nu
,p}~.  \notag
\end{eqnarray}%
Besides, if $g_{rr}=\left( p^{2}+r^{2}\right) ~Y(r)^{-1}$ and $g_{pp}=\left(
p^{2}+r^{2}\right) ~X(p)^{-1}$ one obtains%
\begin{equation*}
R_{rp}=-\frac{1}{2}~\left[ \log (-g~\left( p^{2}+r^{2}\right) ^{3})\right]
_{,rp}-\frac{1}{2~\left( p^{2}+r^{2}\right) }~\left[ p~\left[ \log (-g)%
\right] _{,r}+r~\left[ \log (-g)\right] _{,p}\right] +\frac{1}{4}%
~g_{~~~,r}^{\mu \nu }~g_{\mu \nu ,p}~.
\end{equation*}%
So, the simplest way of changing the Ansatz for the metric, while
keeping the value $R_{rp}=0$, is

i) keep the form of the block $(r,p)$,

ii) modify the block $(t,\phi ,\psi )$ without affecting the values
of $g$ and $g_{~~~,r}^{\mu \nu }~g_{\mu \nu ,p}$.

The preservation of $R_{rp}$ is just one of the clues it should be
observed to have a chance of success in getting a solution to
Einstein-Maxwell equations. Of course, we also should care that this
way of relaxing the Ansatz has not a destructive impact on $F^{\mu
\nu }$; otherwise we would
affect the fulfilling of Maxwell equations, or the form of $T_{~\nu }^{\mu }$%
. Taking these considerations into account, we will extend the Pleba\'{n}%
ski-Demia\'{n}ski Ansatz by replacing~$\mathbf{\omega }^{2}$ in the metric (%
\ref{PD}) with%
\begin{equation}
\mathbf{\Omega }^{2}=\mathbf{\omega }^{2}-\frac{p^{2}~r^{2}}{a~b~\left(
p^{2}+r^{2}\right) }~\left( \mathcal{Y}(r,p)~\mathbf{\omega }^{0}+\mathcal{X}%
(r,p)~\mathbf{\omega }^{1}\right) ~,~  \label{omega2}
\end{equation}%
where $\mathcal{X}(r,p)$, $\mathcal{Y}(r,p)$ are functions to be
chosen. The so-extended Pleba\'{n}ski-Demia\'{n}ski Ansatz is then
\begin{equation}
\mathbf{g}~~=~-\frac{Y(r)}{p^{2}+r^{2}}~~\mathbf{\omega }^{0}\otimes \mathbf{%
\omega }^{0}~+~\frac{X(p)}{p^{2}+r^{2}}~~\mathbf{\omega }^{1}\otimes \mathbf{%
\omega }^{1}~+~\frac{a^{2}~b^{2}}{p^{2}~r^{2}}~\mathbf{\Omega }^{2}\otimes
\mathbf{\Omega }^{2}~+~\left( p^{2}+r^{2}\right) ~\left( \frac{dr\otimes dr}{%
Y(r)}+\frac{dp\otimes dp}{X(p)}\right) ~.  \label{XPD}
\end{equation}%
Notice that the determinant (\ref{det}) is effectively preserved, in spite
of the modifications introduced in the block $(t,\phi ,\psi )$. In fact, the
volume associated with $\mathbf{g}$ is%
\begin{equation}
\text{volume}=\frac{a~b}{~p~r}~\mathbf{\omega }^{0}\wedge \mathbf{\omega }%
^{1}\wedge \mathbf{\Omega }^{2}\wedge dr\wedge dp=\frac{a~b}{~p~r}~\mathbf{%
\omega }^{0}\wedge \mathbf{\omega }^{1}\wedge \mathbf{\omega }^{2}\wedge
dr\wedge dp~;
\end{equation}%
then the volume does not depend on the \textquotedblleft
deformations\textquotedblright\ $\mathcal{X}(r,p)$, $\mathcal{Y}(r,p)$.
Besides we will require that $g_{~~~,r}^{\mu \nu }~g_{\mu \nu ,p}$ keeps its
\textquotedblleft undeformed\textquotedblright\ value, which is%
\begin{equation}
g_{~~~,r}^{\mu \nu }~g_{\mu \nu
,p}~=~-\frac{4}{p~r}-\frac{8~p~r}{\left( p^{2}+r^{2}\right) ^{2}}~;
\label{undeformed_value}
\end{equation}%
this requirement implies that $\mathcal{X}$, $\mathcal{Y}$ must be%
\begin{equation}
\mathcal{X}=h(p)~~~~~~~~\text{and}~~~~~~~~~\mathcal{Y}=q(r)~,~  \label{XXYY1}
\end{equation}%
or%
\begin{equation}
\mathcal{X}=\frac{p^{2}+r^{2}}{p^{2}~r^{2}}\left( p^{2}~h(r)-a~b\right)
~~~~~~~~\text{and}~~~~~~~~~\mathcal{Y}=\frac{p^{2}+r^{2}}{p^{2}~r^{2}}\left(
r^{2}~q(p)-a~b\right) ~.  \label{XXYY2}
\end{equation}%
We will choose the first option because it guarantees the separability of $%
\left( p^{2}+r^{2}\right) ~\mathbf{g}^{-1}$.

It is worth noticing that the CCLP metric \cite{CCLP,DKL} belongs to
this Ansatz; it is the case
\begin{equation}
X_{CCLP}(p)=X_{vac}(p)~,~~~~~~~Y_{CCLP}(r)=Y_{vac}(r)+\frac{Q^{2}+2~a~b~Q}{%
r^{2}}~,~~~~~~~\mathcal{X}_{CCLP}=0~,~~~~~~~\mathcal{Y}_{CCLP}=\frac{Q}{r^{2}%
}~.  \label{CCLP}
\end{equation}%
It could be said that the CCLP metric is somehow \textquotedblleft
biased\textquotedblright\ towards the $Y$, $\mathcal{Y}$ sector.
Hopefully, we might find a solution to Einstein-Maxwell equations by
\textquotedblleft unbiasing\textquotedblright\ the choice of
$\mathcal{X}$, $\mathcal{Y}$.

As can be seen in Eq.~(\ref{XPD}), the basis of $\mathbf{\omega }^{i}$'s~is
no longer orthogonal. Instead the basis $\{\mathbf{\omega }^{0},\mathbf{%
\omega }^{1},\mathbf{\Omega }^{2},dr,dp\}$ is orthogonal in the metric (\ref%
{XPD}). Not surprisingly, the inverse metric also has the Pleba\'{n}ski-Demia%
\'{n}ski structure, but vectors~$\mathbf{v}_{0}$, $\mathbf{v}_{1}$
must be, respectively, substituted for
\begin{equation}
\mathbf{V}_{0}=\mathbf{v}_{0}+\mathcal{Y}(r,p)~\mathbf{v}_{2}~,  \label{V0}
\end{equation}%
\begin{equation}
\mathbf{V}_{1}=\mathbf{v}_{1}+\mathcal{X}(r,p)~\mathbf{v}_{2}~.  \label{V1}
\end{equation}%
The inverse metric reads%
\begin{equation}
\mathbf{g}^{-1}=-\frac{\mathbf{V}_{0}\otimes \mathbf{V}_{0}}{\left(
p^{2}+r^{2}\right) ~Y(r)}~+~\frac{\mathbf{V}_{1}\otimes \mathbf{V}_{1}}{%
\left( p^{2}+r^{2}\right) ~X(p)}~+~\frac{\mathbf{v}_{2}\otimes \mathbf{v}_{2}%
}{p^{2}~r^{2}}~+~\frac{Y(r)}{p^{2}+r^{2}}~\frac{\partial }{\partial r}%
\otimes \frac{\partial }{\partial r}~+~\frac{X(p)}{p^{2}+r^{2}}~\frac{%
\partial }{\partial p}\otimes \frac{\partial }{\partial p}~.
\label{inverseXPD}
\end{equation}%
Thus $\{\mathbf{V}_{0},\mathbf{V}_{1},\mathbf{v}_{2},\partial /\partial
r,\partial /\partial p\}$ is an orthogonal basis in the tangent space that
is dual to $\{\mathbf{\omega }^{0},\mathbf{\omega }^{1},\mathbf{\Omega }%
^{2},dr,dp\}$, except for normalization factors.

We are left with the unknowns $h(p)$, $q(r)$, $X(p)$, and $Y(r)$. If
the extended Pleba\'{n}ski-Demia\'{n}ski Ansatz is useful, we will
success in finding the functions $h$, $q$, $X$, $Y$ endowing the
Einstein tensor with a suitable eigenvalue-eigenvector structure to
be sourced by the energy-momentum tensor belonging to a rotating
\textquotedblleft pointlike\textquotedblright\ charge. So let us now
turn to the electromagnetic potential (\ref{potentialk}), where
$k_{\mu }$ is still the one described in Footnote
\ref{gaugeequivalence}; its contravariant version is the vector
$\mathbf{k}$ in Eq.~(\ref{k}) where $\mathbf{v}_{0}$ must be
replaced with $\mathbf{V}_{0}$. $\mathbf{k}$ is a null and geodesic
vector in the extended metric (\ref{XPD}) too. As was already
pointed out, the
potential (\ref{potentialk}) is gauge equivalent to $\mathbf{A}%
=Q~(p^{2}+r^{2})^{-1}~\mathbf{\omega }^{0}$ and $\mathbf{A}%
=Q~(p^{2}+r^{2})^{-1}~\mathbf{\omega }^{1}$. This equivalence reflects in
the \textquotedblleft unbiased\textquotedblright\ 2-form field%
\begin{equation}
\mathbf{F}=d\mathbf{A}=\frac{2~Q}{(p^{2}+r^{2})^{2}}~\left( r~\mathbf{\omega
}^{0}\wedge dr+p~\mathbf{\omega }^{1}\wedge dp\right) ~.
\end{equation}%
The contravariant field $F^{\mu \nu }$ in the metric (\ref{XPD}) is%
\begin{equation}
F^{\mu \nu }~\frac{\partial }{\partial x^{\mu }}\wedge \frac{\partial }{%
\partial x^{\nu }}=\frac{2~Q}{(p^{2}+r^{2})^{3}}~\left[ r~\mathbf{V}%
^{0}\wedge \frac{\partial }{\partial r}-p~~\mathbf{V}^{1}\wedge \frac{%
\partial }{\partial p}\right] ~.  \label{ExtendedFielduu}
\end{equation}%
The presence of vectors $\mathbf{V}_{0}$, $\mathbf{V}_{1}$ implies
that the extended Ansatz (\ref{XPD}) affects $F^{\mu \nu }$ by
introducing new terms. Even so, the energy-momentum tensor
(\ref{tmunu}) associated to $F^{\mu \nu } $,
\begin{equation}
8\pi G~T_{~\nu }^{\mu }~\frac{\partial }{\partial x^{\mu }}\otimes dx^{\nu
}=-\frac{4~G~Q^{2}}{(p^{2}+r^{2})^{3}}~\left[ \mathbf{V}_{\widehat{0}%
}\otimes \mathbf{\omega }^{\widehat{0}}+\frac{\partial }{\partial r}\otimes
dr-~\mathbf{V}_{\widehat{1}}\otimes \mathbf{\omega }^{\widehat{1}}-\frac{%
\partial }{\partial p}\otimes dp+\frac{p^{2}-r^{2}}{p^{2}+r^{2}}~\mathbf{v}_{%
\widehat{2}}\otimes \mathbf{\Omega }^{\widehat{2}}\right] ~,  \label{Xtmunu}
\end{equation}%
(we are involving the normalized versions (\ref{normal}) of the orthogonal
bases in the tangent and cotangent spaces) exhibits an eigenvalue structure
which does not depend on the choice of $\mathcal{X}$, $\mathcal{Y}$.
Instead, the choice of $\mathcal{X}$, $\mathcal{Y}$ does affect the
fulfillment of Maxwell equations. However, by choosing the functions $h(p)$,
$q(r)$ in Eq.~(\ref{XXYY1}) as
\begin{equation}
\mathcal{X}(r,p)=\frac{\mu _{\mathcal{X}}~Q}{p^{2}}+\zeta ~,~~~~~~~~~~~~~~~%
\mathcal{Y}(r,p)=\frac{\mu _{\mathcal{Y}}~Q}{r^{2}}-\zeta ~,
\label{ExtendedXXYY}
\end{equation}%
we obtain that $F^{\mu \nu }$ fulfills Maxwell-Chern-Simons equations,%
\begin{equation}
\partial _{\mu }\left( \sqrt{-g}~F^{\mu \nu }\right) =\frac{1}{4}~\left( \mu
_{\mathcal{X}}-\mu _{\mathcal{Y}}\right) ~\epsilon ^{\nu \lambda \rho \alpha
\beta }~F_{\lambda \rho }~F_{\alpha \beta }~,  \label{MCS}
\end{equation}%
with an arbitrary coupling constant proportional to $\mu _{\mathcal{X}}-\mu
_{\mathcal{Y}}$ (the value of $\zeta $ is irrelevant, not only at this stage
but for the rest of the analysis as well.).\footnote{%
Even if we work with Maxwell-Chern-Simons equations, the Maxwellian form of $%
T_{~\nu }^{\mu }$ remains valid. This is because the Chern-Simons
term in the action, $F\wedge F\wedge A$, does not contain the
metric; thus it does not contribute to $T_{~\nu }^{\mu
}$.\label{CS}}

The potential (\ref{potentialk}) satisfies Eq.~(\ref{MCS}) whatever
the functions $X(p)$ and $Y(r)$ are. So we are now left with two
unknowns, $X(p)$ and $Y(r)$. Our aim is to properly choose them to
obtain solutions to Einstein-Chern-Simons equations with arbitrary
Chern-Simons coupling constant.\bigskip\

Concerning the structure of $G_{~~\nu }^{\mu }+\Lambda ~\delta _{~\nu }^{\mu
}$ in the extended Pleba\'{n}ski-Demia\'{n}ski Ansatz, one can verify that $%
\partial /\partial r$ and $\partial /\partial p$ are still eigenvectors
whatever $X$, $Y\,\ $are. As a necessary condition to match the
energy-momentum tensor (\ref{Xtmunu}), their respective eigenvalues $%
\epsilon _{3}$, $\epsilon _{4}$ should be equal and opposite.
Without loss of generality, let us we write
\begin{equation}
X(p)=X_{vac}(p)+f(p)~,
\end{equation}%
\begin{equation}
Y(r)=Y_{vac}(r)+g(r)~;
\end{equation}%
then one gets that the eigenvalues are effectively equal and opposite if and
only if%
\begin{equation}
p^{-3}~(p^{3}~f^{\prime })^{\prime }+r^{-3}~(r^{3}~g^{\prime })^{\prime }=0~.
\label{fg}
\end{equation}%
This linear equation has independent inverse square homogeneous solutions
for $f$ and $g$,\footnote{%
Equation (\ref{fg}) also accepts the solution $f=-\beta
~p^{2}+\gamma $, $g=\beta
~r^{2}+\delta $; however this solution is already present in $X_{vac}$, $%
Y_{vac}$ through the integration constants $\alpha $, $m$, $n$.} which
implies two new constants in $X$, $Y$. For convenience, we will write the
solutions in the following way:%
\begin{equation}
X(p)=X_{vac}(p)-\mu _{\mathcal{X}}~Q~\frac{\mu _{\mathcal{X}}~Q+2~a~b+\beta
_{X}}{p^{2}}~,
\end{equation}%
\begin{equation}
Y(r)=Y_{vac}(r)+\mu _{\mathcal{Y}}~Q~\frac{\mu _{\mathcal{Y}}~Q+2~a~b+\beta
_{Y}}{r^{2}}~.
\end{equation}%
So far, the results for $X$, $Y$ do not differ from the ones we
would have obtained in the previous Section. However we have changed
the geometry, by extending the Pleba\'{n}ski-Demia\'{n}ski Ansatz,
as an attempt to match the eigenvalues and eigenvectors of $G_{~~\nu
}^{\mu }+\Lambda ~\delta _{~\nu }^{\mu }$ with those of the source
(\ref{Xtmunu}). In fact, the extended
Ansatz replaced vectors $\mathbf{v}_{0},\mathbf{v}_{1}$ with $\mathbf{V}_{0},%
\mathbf{V}_{1}$ as defined in Eqs.~(\ref{V0}), (\ref{V1}), (\ref%
{ExtendedXXYY}). Thus we obtain that not only $\partial /\partial r,\partial
/\partial p$ but $\mathbf{V}_{0},\mathbf{V}_{1},\mathbf{v}_{2}$ are
eigenvectors of $G_{~~\nu }^{\mu }+\Lambda ~\delta _{~\nu }^{\mu }$ with
eigenvalues%
\begin{eqnarray}
-\epsilon _{0} &=&-\epsilon _{3}=~\epsilon _{1}=\epsilon _{4}=~\frac{%
3~Q^{2}~\left( \mu _{\mathcal{X}}-\mu _{\mathcal{Y}}\right) ^{2}}{%
(p^{2}+r^{2})^{3}}+Q~\frac{(3~p^{2}+r^{2})~r^{4}~\mu _{\mathcal{X}}~\beta
_{X}+(p^{2}+3~r^{2})~p^{4}~\mu _{\mathcal{Y}}~\beta _{Y}}{%
p^{4}~r^{4}~(p^{2}+r^{2})^{2}}~,~~~~ \\
\epsilon _{2} &=&-3~Q^{2}~\left( \mu _{\mathcal{X}}-\mu _{\mathcal{Y}%
}\right) ^{2}~\frac{p^{2}-r^{2}}{(p^{2}+r^{2})^{4}}+3~Q~\frac{p^{4}~\mu _{%
\mathcal{Y}}~\beta _{Y}-~r^{4}~\mu _{\mathcal{X}}~\beta _{X}}{%
p^{4}~r^{4}~(p^{2}+r^{2})}~.
\end{eqnarray}%
The terms proportional to $Q^{2}\left( \mu _{\mathcal{X}}-\mu _{\mathcal{Y}%
}\right) ^{2}$ are characteristic of this extended Ansatz. So, we
recognize at least two interesting cases:

\bigskip

i) If $\beta _{X}=0=\beta _{Y}$, then the eigenvalues of $G_{~~\nu }^{\mu
}+\Lambda ~\delta _{~\nu }^{\mu }$ coincide with those of $8\pi G~T_{~\nu
}^{\mu }$ (i.e., Einstein equations are verified) provided that%
\begin{equation}
\left( \mu _{\mathcal{X}}-\mu _{\mathcal{Y}}\right) ^{2}=\frac{4~G}{3}~~.
\label{condition}
\end{equation}%
Thus the rotating \textquotedblleft pointlike\textquotedblright\
charged solution to Einstein-Maxwell-Chern-Simons equations is
obtained only for a
specific value of the Chern-Simons coupling constant. The CCLP solution (\ref%
{CCLP}) is a particular choice of $\mu _{\mathcal{X}}$, $\mu _{\mathcal{Y}}$
satisfying the condition (\ref{condition}). However, each choice of\ $\mu _{%
\mathcal{X}}$, $\mu _{\mathcal{Y}}$ fulfilling Eq.~(\ref{condition})
still could imply a different geometry. In fact the Kretschmann
invariant
depends not only on $\mu _{\mathcal{X}}-\mu _{\mathcal{Y}}$ but on $\mu _{%
\mathcal{X}}$, $\mu _{\mathcal{Y}}$ in a separate way. For instance, in the
simplest case $\lambda =m=n=0$ it results
\begin{eqnarray}
R_{~~\lambda \rho }^{\mu \nu }~R_{~~\mu \nu }^{\lambda \rho }\ &=&\ \frac{%
4~Q^{2}\left( \mu _{\mathcal{X}}-\mu _{\mathcal{Y}}\right) ^{2}}{%
(p^{2}+r^{2})^{8}}\ \Bigg[\frac{48~(a^{2}+b^{2})~(p^{4}-r^{4})^{2}}{%
p^{2}-r^{2}}  \notag \\
&&  \notag \\
&&+192~\left( a~b~(p^{2}+r^{2})+Q~(p^{2}\mu _{\mathcal{X}}+r^{2}\mu _{%
\mathcal{Y}})\right) ^{2}+Q^{2}\left( \mu _{\mathcal{X}}-\mu _{\mathcal{Y}%
}\right) ^{2}~\left( 28~p^{2}r^{2}-65~(p^{4}+r^{4})\right) \Bigg]\ .
\label{Kretschmann}
\end{eqnarray}%
So different values of $\mu _{\mathcal{X}}$, $\mu _{\mathcal{Y}}$,
subjected to the condition (\ref{condition}), lead to different
values of the Kretschmann invariant. This is a good indication to
think that we have found a one-parametric family of geometries.
However, the certainty must come from a proper global analysis of
the involved solutions.

\bigskip

ii) If $\beta _{X}=\beta _{Y}$ and $\mu _{\mathcal{X}}=\mu _{\mathcal{Y}}$,
then the eigenvalues become%
\begin{eqnarray}
-\epsilon _{0} &=&-\epsilon _{3}=~\epsilon _{1}=\epsilon _{4}=Q~~\mu _{%
\mathcal{X}}~\beta _{X}~\frac{p^{2}+r^{2}}{p^{4}~r^{4}}~,~~~~ \\
\epsilon _{2} &=&3~Q~\mu _{\mathcal{X}}~\beta _{X}~\frac{p^{2}-r^{2}~}{%
p^{4}~r^{4}}~.
\end{eqnarray}%
It would be interesting to look for sources matching this eigenvalue
structure.

\section{Further Extensions of Pleba\'{n}ski-Demia\'{n}ski Ansatz}

\label{further extensions}

Once we have understood the mechanism to make the CCLP-like
solutions work, we can try extending this mechanism to search for
more solutions. For instance, we could further extent the Ansatz
(\ref{omega2}) by considering that $\mathbf{\Omega }^{2}$ could have
also components $dr$ and $dp$:
\begin{equation}
\mathbf{\Omega }^{2}=\mathbf{\omega }^{2}-\frac{p^{2}~r^{2}}{a~b~\left(
p^{2}+r^{2}\right) }~\left( \mathcal{Y}(r,p)~\mathbf{\omega }^{0}+\mathcal{X}%
(r,p)~\mathbf{\omega }^{1}\right) -\frac{p^{2}~r^{2}}{a~b}~\left( \mathcal{Z}%
(r,p)~dr+\mathcal{W}(r,p)~dp\right) ~,~~
\end{equation}%
which not only forces the replacements (\ref{V0}) and (\ref{V1}) in the
inverse metric, but the replacement of vectors $\partial /\partial r$ and $%
\partial /\partial p$ with%
\begin{equation}
\mathbf{V}_{3}=\frac{\partial }{\partial r}+\mathcal{Z}(r,p)~\mathbf{v}_{2}~,
\label{V3}
\end{equation}%
\begin{equation}
\mathbf{V}_{4}=\frac{\partial }{\partial p}+\mathcal{W}(r,p)~\mathbf{v}_{2}~,
\label{V4}
\end{equation}%
Thus the inverse metric becomes%
\begin{equation}
\mathbf{g}^{-1}=-\frac{\mathbf{V}_{0}\otimes \mathbf{V}_{0}}{\left(
p^{2}+r^{2}\right) ~Y(r)}~+~\frac{\mathbf{V}_{1}\otimes \mathbf{V}_{1}}{%
\left( p^{2}+r^{2}\right) ~X(p)}~+~\frac{\mathbf{v}_{2}\otimes \mathbf{v}_{2}%
}{p^{2}~r^{2}}~+~\frac{Y(r)}{p^{2}+r^{2}}~\mathbf{V}_{3}\otimes \mathbf{V}%
_{3}~+~\frac{X(p)}{p^{2}+r^{2}}~\mathbf{V}_{4}\otimes \mathbf{V}_{4}~.
\label{XXIPD}
\end{equation}%
The determinant of the metric remains independent of $X$, $Y$, $\mathcal{X}$%
, $\mathcal{Y}$, $\mathcal{W}$, $\mathcal{Z}$. The potential (\ref%
{potentialk}) still satisfies Maxwell-Chern-Simons equations (\ref{MCS}) for
the choices (\ref{ExtendedXXYY}), and its energy-momentum tensor is
\begin{equation}
8\pi G~T_{~\nu }^{\mu }~\frac{\partial }{\partial x^{\mu }}\otimes dx^{\nu
}=-\frac{4~G~Q^{2}}{(p^{2}+r^{2})^{3}}~\left[ \mathbf{V}_{\widehat{0}%
}\otimes \mathbf{\omega }^{\widehat{0}}+\mathbf{V}_{3}\otimes dr-~\mathbf{V}%
_{\widehat{1}}\otimes \mathbf{\omega }^{\widehat{1}}-\mathbf{V}_{4}\otimes
dp+\frac{p^{2}-r^{2}}{p^{2}+r^{2}}~\mathbf{v}_{\widehat{2}}\otimes \mathbf{%
\Omega }^{\widehat{2}}\right] ~,  \label{XXTmunu}
\end{equation}%
The separability of $\left( p^{2}+r^{2}\right) ~\mathbf{g}^{-1}$ is
guaranteed by choosing%
\begin{equation}
\mathcal{W}(r,p)=w(p)~,~~~~~~~~~~~~~~~\mathcal{Z}(r,p)=z(r)~.  \label{WZ}
\end{equation}%
Although the metric is no longer diagonal by blocks in the chart $(t,\phi
,\psi ,r,p)$, $R_{~r}^{p}$ is still zero. However the Ricci tensor $R_{~\nu
}^{\mu }$becomes non-linear in $X$, $Y$; its form is complicated enough to
suggest that this strategy will not be successful in getting solutions to
Einstein-Maxwell-Chern-Simons equations for other values of the Chern-Simons
coupling constant (however, see Ref.~\cite{AC} for the slowly rotating case).

As a different try, we will invert the procedure of the previous section.
Instead of changing $\mathbf{\omega }^{2}$ in the metric (\ref{PD}), we will
change $\mathbf{v}_{2}$ in the inverse metric (\ref{IPD}). This is a rather
obvious strategy, once one realizes that $\mathbf{g}$ and $\mathbf{g}^{-1}$
are on an equal footing in the requirements to keep $R_{rp}=0$. If $\mathbf{v%
}_{2}$ is substituted with%
\begin{equation}
\mathbf{V}_{2}=\mathbf{v}_{2}-\frac{p^{2}~r^{2}}{a~b~\left(
p^{2}+r^{2}\right) }~\left( \mathcal{N}(r)~\mathbf{v}_{0}+\mathcal{M}(p)~%
\mathbf{v}_{1}\right) ~,
\end{equation}%
then the determinant of $\mathbf{g}^{-1}$ will be preserved; however the
separability of $(p^{2}+r^{2})~\mathbf{g}^{-1}$ will be lost. The
corresponding metric $\mathbf{g}$ can be reached by replacing in (\ref{PD}):%
\begin{eqnarray}
\mathbf{\Omega }^{0} &=&\mathbf{\omega }^{0}+~\mathcal{N}(r)~\mathbf{\omega }%
^{2}~, \\
&& \notag \\
\mathbf{\Omega }^{1} &=&\mathbf{\omega }^{1}+~\mathcal{M}(p)~\mathbf{\omega }%
^{2}~,
\end{eqnarray}%
By keeping the ``undeformed'' value (\ref{undeformed_value}) we are led to%
\begin{equation}
\mathcal{N}(r)=\frac{1}{1+\mu ~r^{2}}~,~~~~~~~\mathcal{M}(p)=0~,
\end{equation}%
or%
\begin{equation}
\mathcal{N}(r)=0~,~~~~~~~\mathcal{M}(p)=\frac{1}{1+\mu ~p^{2}}~.
\end{equation}%
Nevertheless, the potential (\ref{potentialk}) is not a solution to
Maxwell-Chern-Simons equations in this geometry.

\section{Pure-radiation solution to $5D$ Einstein-Maxwell equations}

\label{Pure radiation}

We will leave the Pleba\'{n}ski-Demia\'{n}ski family of metrics, to
show a different type of non-static solutions to $5D$
Einstein-Maxwell equations in the framework of the Kerr-Schild
Ansatz (\ref{KSansatz}). We will still use
the Pleba\'{n}ski-Demia\'{n}ski form to introduce a suitable seed metric $%
\overset{o}{g}_{\mu \nu }$. In fact, we will start from the metric (\ref{PD}%
), as written for the vacuum solutions (\ref{Xvac}) and (\ref{Yvac}) with
the following choice of constants:%
\begin{equation}
b=0~,~~~~~~~~\alpha =1-a^{2}~\lambda ~,~~~~~~~~m=\frac{a^{2}}{2}=-n~.
\label{choice}
\end{equation}%
Therefore the functions $X_{vac}$, $Y_{vac}$ become%
\begin{equation}
X_{vac}(p)=-\lambda ~p^{4}~,~~~~~~~~~~~~~~Y_{vac}(r)=-\lambda ~r^{4}~.
\label{AdSXY}
\end{equation}%
Since both $X_{vac}$, $Y_{vac}$ must be positive in order to have a
Lorentzian metric, then one gets that $\lambda $ has to be negative; besides
it is $m=-n$, what means that the seed metric is the anti-de Sitter ($AdS$)
geometry in a peculiar chart: $\overset{o}{g}_{\mu \nu }=\mathbf{g}_{AdS}$.
In such chart, the interval associated with the seed metric is%
\begin{equation}
ds_{AdS}^{2}~=~(p^{2}-r^{2})~du^{2}\,+p^{2}r^{2}~(2~du~d\sigma
+dw^{2})-\lambda ^{-1}\,\left( p^{2}+r^{2}\right) \,\left( \frac{dr^{2}}{%
r^{4}}+\frac{dp^{2}}{p^{4}}\right) ~,  \label{AdS}
\end{equation}%
where%
\begin{equation}
u=\frac{\sqrt{-\lambda }~(t-a~\phi )}{1-a^{2}~\lambda }~,~~~~~~~~~~~\
~~~\sigma =a^{-2}~\left( u-\sqrt{-\lambda }~t\right) ~,~~~~~~~~~~~\ ~~~w=%
\frac{\psi }{a}~.~
\end{equation}%
As can be seen, the parameter $a$ has been absorbed into the new
coordinates; no trace of $a$ remains in the interval (\ref{AdS}). With the
help of the complex coordinate%
\begin{equation}
\zeta =\frac{1}{2~\sqrt{-\lambda }}~\left( \frac{1}{r}+\frac{i}{p}\right)
^{2}  \label{complex}
\end{equation}%
the interval reads%
\begin{equation}
ds_{AdS}^{2}~=~4~\lambda ^{-1}(\zeta -\overline{\zeta })^{-2}~\left[
du~\left( \sqrt{-\lambda }~(\zeta +\overline{\zeta })~du+2~d\sigma \right)
+dw^{2}+~d\zeta ~d\overline{\zeta }\right] \,\,~,
\end{equation}%
or using the real and imaginary parts of $\zeta =\chi +i~y$,
\begin{equation}
ds_{AdS}^{2}~=~-\lambda ^{-1}y^{-2}~\left[ du~\left( 2\sqrt{-\lambda }~\chi
~du+2~d\sigma \right) +dw^{2}+~d\chi ^{2}+dy^{2}\right] \,\,~.  \label{AdS2}
\end{equation}%
The metric inside the square brackets is flat. In fact, in the chart $(\tau
,w,x,y,z)$ such that
\begin{equation}
u=\tau +z~,~~~~~~~~~~~\sigma ~=\frac{\lambda ~(\tau +z)^{3}}{3}-\frac{(\tau
-z)}{2}-\sqrt{-\lambda }~(\tau +z)~x~,\ ~~~~~~~~~~\chi ~=x+~\frac{\sqrt{%
-\lambda }~(\tau +z)^{2}}{2}~~,
\end{equation}%
the interval reads%
\begin{equation}
ds_{AdS}^{2}~=~-\lambda ^{-1}y^{-2}~\left[ -d\tau
^{2}+dw^{2}+dx^{2}+dy^{2}+dz^{2}\right] \,~,
\end{equation}%
which is one of known forms the $AdS$ metric can adopt (see Sec. 5.3
in Ref. \cite{GP}).

Coming back to the chart $(u,\sigma ,w,r,p)$, let us introduce the
electromagnetic potential%
\begin{equation}
\mathbf{A}=\frac{A(u)}{~p^{2}~r^{2}}~\mathbf{n~,}  \label{potentialKS}
\end{equation}%
where $\mathbf{n}\equiv du$ is the null 1-form of components%
\begin{equation}
n_{\mu }=\{1,~0,~0,~0,~0\}~.  \label{vectorn}
\end{equation}%
The electromagnetic potential (\ref{potentialKS}) fulfills the Maxwell
equations not only in the geometry (\ref{AdS}) but in any metric having the
Kerr-Schild form%
\begin{equation}
\mathbf{g}\ =\ \mathbf{g}_{AdS}+f(u,r,p)\,\mathbf{n}\,\otimes \mathbf{n}~.
\label{metric-pure-radiation}
\end{equation}%
This solution is a pure-radiation field, since $T_{~\nu }^{\mu }$
has the
form\footnote{%
The general solution for Maxwell equations in the considered metric (\ref%
{AdS}), (\ref{metric-pure-radiation}) is $\mathbf{A}_{\kappa
}=p^{-1}r^{-1}~\left( G_{\kappa }(u)~J_{1}(\kappa ~p^{-1})+H_{\kappa
}(u)~Y_{1}(\kappa ~p^{-1})\right) $ $~\left( P_{\kappa }(u)~I_{1}(\kappa
~r^{-1})+Q_{\kappa }(u)~K_{1}(\kappa ~r^{-1})\right) ~\mathbf{n}$, but it
does not guarantee the pure-radiation form of the energy-momentum tensor.}%
\begin{equation}
T_{~\nu }^{\mu }=\frac{-\lambda ~A(u)^{2}}{\pi ~p^{4}~r^{4}}~n^{\mu }~n_{\nu
}~.  \label{pure-radiation}
\end{equation}%
On the other hand, the Einstein tensor for the metric (\ref%
{metric-pure-radiation}) is%
\begin{equation}
G_{~\nu }^{\mu }+\Lambda ~\delta _{~\nu }^{\mu }=-2~\lambda \left( ~f-\frac{1%
}{4~(p^{2}+r^{2})}~\left[ p^{3}~\frac{\partial }{\partial p}\left( p~\frac{%
\partial f}{\partial p}\right) +r^{3}~\frac{\partial }{\partial r}\left( r~%
\frac{\partial f}{\partial r}\right) \right] \right) ~n^{\mu }~n_{\nu }~.
\label{Einstein-pure-radiation}
\end{equation}%
The fact that both the energy-momentum tensor (\ref{pure-radiation})
and the Einstein tensor (\ref{Einstein-pure-radiation}) have the
same structure implies that the Kerr-Schild Ansatz is successful in
this case, because there is a sole equation to be satisfied by the
unknown function $f(u,r,p)$.
In fact, Einstein-Maxwell equations are satisfied by choosing\footnote{%
Actually Einstein-Maxwell-Chern-Simons equations are fulfilled too. On the
one hand, the Chern-Simons term $\epsilon ^{\nu \lambda \rho \alpha \beta
}~F_{\lambda \rho }~F_{\alpha \beta }$ vanishes for potentials like (\ref%
{potentialKS}), since the only non-null independent components of
the field tensor are $F_{ur}$ and $F_{up}$. On the other hand, the
Chern-Simons coupling does not contribute to the energy-momentum
tensor, as mentioned in Footnote \ref{CS}.}
\begin{equation}
f(u,r,p)=-\frac{2}{7}~G~A(u)^{2}~\frac{(p^{2}+r^{2})^{2}}{p^{6}~r^{6}}~~.
\label{fA}
\end{equation}%
Besides, the function $f$ can be added with a homogeneous solution of
Einstein equations, like
\begin{equation}
f_{vac}=\frac{\beta (u)}{p^{2}~r^{2}}+\frac{\gamma (u)~p^{8}~r^{8}}{%
(p^{2}+r^{2})^{5}}+\delta (u)~p^{2}~r^{2}+\kappa (u)~\frac{(p^{2}+r^{2})^{3}%
}{p^{4}~r^{4}}+\frac{\varepsilon (u)}{p^{4}~r^{4}}~\left( \frac{%
3~(p^{2}+r^{2})^{2}}{14~p^{2}~r^{2}}-1\right)  \label{fvac}
\end{equation}%
which generates vacuum solutions representing exact gravitational waves
associated with the null 1-form $n_{\mu }$.\footnote{%
The term $\delta (u)~p^{2}~r^{2}$ should be absorbable by changing
coordinates, since it does not take part in the Riemann tensor.} Actually
the (linear) homogeneous equation for $f_{vac}$ can be solved by separating
variables; the general solution is obtained by linearly combining the
solutions%
\begin{equation}
f_{vac}~_{\nu }=\left( C_{\nu }(u)~J_{2}(\nu ~p^{-1})+D_{\nu }(u)~Y_{2}(\nu
~p^{-1})\right) ~\left( E_{\nu }(u)~I_{2}(\nu ~r^{-1})+F_{\nu }(u)~K_{2}(\nu
~r^{-1})\right) ~,  \label{fvacBessel}
\end{equation}%
where $C_{\nu }$, $D_{\nu }$, $E_{\nu }$, $F_{\nu }$ are arbitrary functions
of $u$, and $\nu $ is a separation constant.

In sum, we have obtained an electrovacuum solution whose metric is conformal
to a $pp$wave in $5D$. In fact, according to Eqs.~(\ref{AdS2}) and (\ref%
{metric-pure-radiation}) the interval is
\begin{equation}
ds^{2}~=~-\lambda ^{-1}y^{-2}~\left[ du~\left( (2\sqrt{-\lambda }~\chi
-\lambda ~y^{2}~f~)~du+2~d\sigma \right) +dw^{2}+~d\chi ^{2}+dy^{2}\right]
\,\,~,  \label{pp}
\end{equation}%
which belongs to the class of metrics studied by Kundt \cite{K}, and has the
form of Siklos metric \cite{S} (see also Refs. \cite{WT}, \cite{MTT}).
However the functions (\ref{fA}-\ref{fvacBessel}) are characteristic of five
dimensions. Concerning the $pp$wave metric inside the bracket of Eq.~(\ref%
{pp}), the 1-form (\ref{vectorn}) has zero covariant derivative whatever the
function $f $ is, and is a null vector of the respective Weyl tensor.

\bigskip

\section{Conclusions}

\label{Conclusions}

Pleba\'{n}ski-Demia\'{n}ski-like Ansatz (\ref{PD}) does not contain
the solution to Einstein-Maxwell equations for a rotating
\textquotedblleft pointlike\textquotedblright\ charge in five
dimensions. The reason can be traced to the lack of agreement
between the eigenvalues of the
energy-momentum tensor and those belonging to the Einstein tensor (\ref%
{source}) in such an Ansatz, even though the eigenvector structures
do coincide. However, the Pleba\'{n}ski-Demia\'{n}ski Ansatz can be
properly extended to obtain a wider framework where the
eigenvector-eigenvalue structure of the
energy-momentum and Einstein tensors can be matched, as shown in Section \ref%
{extendedPD}. In this extended Ansatz the CCLP solution is found.
Remarkably, the CCLP geometry is just one in a set of solutions
satisfying
Einstein-Maxwell-Chern-Simons equations for the electromagnetic potential (%
\ref{potentialk}) and the coupling constant $\mu _{\mathcal{X}}-\mu _{%
\mathcal{Y}}=2\sqrt{G/3}$ in Eq.~(\ref{MCS}). In fact, this constraint on $%
\mu _{\mathcal{X}}$, $\mu _{\mathcal{Y}}$ still leaves alive many different
solutions fulfilling the equations, as evidenced by the fact that the
Kretschmann invariant in Eq.~(\ref{Kretschmann}) depends independently on $%
\mu _{\mathcal{X}}$ and $\mu _{\mathcal{Y}}$. Whether they are different
geometries or not should be elucidated through the analysis of the global
properties of these solutions.

Although we tried other extensions of Pleba\'{n}ski-Demia\'{n}ski
Ansatz, we did not succeed in getting solutions sourced by a
rotating pointlike charge for other values of the Chern-Simons
coupling constant. Instead we have obtained a family of
gravitational waves depending on three coordinates, with and without
a pure-radiation electromagnetic field. These geometries are
described by the interval (\ref{pp}), where $f$ can be substituted
with
a combination of the functions (\ref{fA}), (\ref{fvac}), and (\ref%
{fvacBessel}).

\bigskip

\begin{acknowledgments}
The author is indebted to Nathalie Deruelle for proposing and
encouraging this research, and the Universit\'{e} Paris Diderot for its
hospitality and the financial support granted by the {\it Chaire
Alicia Moreau}. The author also thanks A.~Anabal\'{o}n, M.~C\'{a}rdenas,
O.~Fuentealba, G.~Giribet, and F.~Juli\'{e} for helpful comments.
\end{acknowledgments}

\end{document}